\documentclass[10pt,a4paper]{article}

\usepackage[
    a4paper,
    top=18mm,
    bottom=18mm,
    left=22mm,
    right=22mm
]{geometry}

\usepackage[T1]{fontenc}
\usepackage{amsmath}
\usepackage{amssymb}
\usepackage{mathtools}
\usepackage{amsthm}
\usepackage{XCharter}
\usepackage[xcharter]{newtxmath}
\usepackage{microtype}

\usepackage{booktabs}
\usepackage{tabularx}
\usepackage{array}
\usepackage{multirow}
\usepackage{makecell}
\usepackage{graphicx}
\usepackage{grffile}
\usepackage[table]{xcolor}
\usepackage{enumitem}
\usepackage[normalem]{ulem}
\usepackage[section]{placeins}
\usepackage{float}
\usepackage{caption}
\usepackage{titlesec}
\usepackage{fancyhdr}
\usepackage{tikz}
\usepackage{pgfplots}
\pgfplotsset{compat=1.18}
\usepackage{algorithm}
\usepackage[noend]{algpseudocode}
\usepackage{flafter}
\usepackage{xurl}
\usepackage[square,authoryear]{natbib}
\let\cite\citep
\usepackage{hyperref}

\usetikzlibrary{
    shapes.geometric,
    arrows.meta,
    positioning,
    calc,
    fit,
    backgrounds,
    decorations.pathreplacing
}

\definecolor{secondblue}{RGB}{28,94,190}
\definecolor{fieldblue}{RGB}{221,235,246}
\definecolor{processgreen}{RGB}{227,239,223}
\definecolor{bankorange}{RGB}{250,234,214}
\definecolor{latentpurple}{RGB}{235,226,244}
\definecolor{linkblue}{RGB}{0,60,180}

\hypersetup{
    colorlinks=true,
    linkcolor=linkblue,
    citecolor=linkblue,
    urlcolor=linkblue,
    pdfauthor={Mahipal Jetta and Sujato Dutta},
    pdftitle={Hybrid++: The Bridge between PDE Models and Deep Learning for Gamma Noise Removal}
}

\titleformat{\section}
    {\large\bfseries}
    {\thesection.}{0.55em}{}
\titleformat{\subsection}
    {\normalsize\bfseries}
    {\thesubsection.}{0.50em}{}
\titleformat{\subsubsection}
    {\normalsize\bfseries}
    {\thesubsubsection.}{0.50em}{}
\titlespacing*{\section}{0pt}{1.35em}{0.55em}
\titlespacing*{\subsection}{0pt}{1.05em}{0.35em}
\titlespacing*{\subsubsection}{0pt}{0.90em}{0.30em}

\DeclareCaptionLabelSeparator{verticalbar}{\enspace|\enspace}
\newcommand{\shorttitle}{Hybrid++: PDE Models and Deep Learning for Gamma Noise Removal}
\renewcommand{\arraystretch}{1.09}

\begin{document}
\thispagestyle{empty}
\vspace*{-7mm}

\noindent
\IfFileExists{mu_logo.png}{%
    \includegraphics[height=24mm,keepaspectratio]{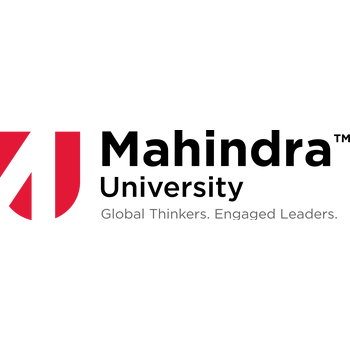}%
}{%
    {\fontsize{14}{18}\selectfont\bfseries MAHINDRA UNIVERSITY}%
}
\par\vspace{-4mm}
\noindent\rule{\linewidth}{0.55pt}
\vspace{6mm}

{\fontsize{23}{27}\selectfont\bfseries
Hybrid++: The Bridge between\\
PDE Models and Deep Learning\\
for Gamma Noise Removal
\par}
\vspace{6mm}

{\normalsize
Mahipal Jetta,\quad Sujato Dutta
\par}
\vspace{1.5mm}
{\small
Mahindra University
\par}
\vspace{7mm}

\noindent{\large\bfseries Abstract}\par\vspace{2mm}
\begingroup\small\noindent
\textbf{Multiplicative gamma noise is one of the dominant noise factors in Synthetic Aperture Radar (SAR) and medical ultrasound images. They are dependent on pixel level noises due to which they are highly varying across the image and harder to handle as compared to additive noise. The denoising methods to address this noise currently include classical partial differential equation (PDE) methods which are good for interpretation but lack the restoration ability as compared to the state-of-the-art models while deep neural networks like DnCNN achieve high performance but at the cost of transparency due to which practitioners are skeptical to use them in high-risk critical fields like medicine. This paper presents Hybrid++, a novel trainable nonlinear reaction-diffusion architecture that addresses both the concerns - staying interpretable while offering performance close to huge black-box models. It combines a fully learnable PDE initialization with a 3-stage reaction-diffusion network having 64-channel multiscale filter banks, 4-layer Squeeze-and-Excitation attention-based influence functions and a 64-dimensional noise level embedding. It uses a two-phase training strategy, stage-wise optimization followed by joint end-to-end refinement which enables co-adaptation of all learnable parameters. On the FoE benchmark, Hybrid++ substantially improves over classical PDE, BM3D and the original TNRD baselines. In the severe-noise setting L=1, it comes within 0.23 dB PSNR of a separately trained DnCNN while using only about 8\% of its parameters.}

\par\endgroup
\vspace{2mm}
\noindent{\small\textbf{Keywords:} Gamma noise; Image Denoising; Partial Differential Equations (PDE).}\par
\vspace{2mm}

\section{Introduction}
\label{sec:intro}
Multiplicative gamma noise is a significant challenge to deal with in imaging systems such as Synthetic Aperture Radar (SAR) and medical ultrasound where the noise is signal-dependent and cannot be handled by standard additive-noise denoising models. Classical methods like partial differential equations (PDEs) \cite{perona1990} try to address this by adapting smoothing strength to local image gradients and intensity. However, they rely on fixed, handcrafted parameters and struggle to recover complex features from the noisy images. On the other hand, deep convolutional neural networks (CNNs) such as DnCNN \cite{Zhang_2017} achieve remarkable performance in denoising but operate as black box models which can't be easily interpreted and lack transparency. They also require training hundreds of thousands of parameters. Traditional non-local methods like BM3D \cite{dabov2007bm3d}, are effective for additive Gaussian noise but require ad-hoc adaptations (e.g. homomorphic transforms) for multiplicative noise and often fail to produce optimal results.
Trainable Nonlinear Reaction Diffusion (TNRD) \cite{Chen_2017} was one of the breakthroughs offering a  middle ground by unfolding the denoising process as a sequence of interpretable diffusion stages with learnable parameters. However, the original TNRD was meant mainly for additive noise and its compact parameterization limits its capacity for handling the more challenging multiplicative noise. Recent restoration models explore efficient CNNs, hybrid transformers and self-supervised attention, as exemplified by NAFNet, Xformer and SwinIA \cite{nafnet2022,xformer2024,swinia2025}. These advances motivate examining both restoration accuracy and model complexity. 

We design \textbf{Hybrid++} as a compact, interpretable extension of the PDE/reaction-diffusion family for multiplicative gamma noise. Section~\ref{sec:related} places this design in the broader recent literature.
Our main contributions are:
\begin{enumerate}
    \item A fully differentiable PDE initialization module whose parameters ($\alpha$, $\beta$) are optimized via backpropagation in place of traditional grid search.
    \item Parallel multiscale filter banks (32 filters of $3{\times}3$ and 32 filters of $5{\times}5$) producing a 64-channel feature representation that captures both fine textures and gradients while also handling broader structural context.
    \item A 4-layer attention influence function with an expanded squeeze-excitation \cite{hu2019squeezeandexcitationnetworks} bottleneck (64$\to$16$\to$64) (here the numbers denote the number of neurons in the respective layers of the network), enabling deep nonlinear feature weighting without normalization layers.
    \item A 3-layer MLP \cite{popescu2009mlp} noise embedding (1$\to$32$\to$64$\to$64) that conditions every stage on the noise severity allowing a single unified model to handle varying noise levels.
\end{enumerate}

\section{Related Work}
\label{sec:related}
Recent image restoration spans convolutional, attention-based and generative models. NAFNet \cite{nafnet2022} simplifies a convolutional restoration backbone using multiplicative gating while MAXIM \cite{maxim2022} uses multi-axis gated MLPs to combine local and global interactions.

Transformers provide a different route to contextual modeling. SwinIR \cite{liang2021swinirimagerestorationusing} uses shifted-window attention, Restormer \cite{zamir2022restormerefficienttransformerhighresolution} uses efficient transposed attention, and Uformer \cite{Wang_2022_Uformer} combines window attention with a U-shaped multiscale hierarchy. Xformer \cite{xformer2024} combines spatial and channel interactions in a hybrid X-shaped denoising architecture. More recent state-space models, MambaIR \cite{mambair2024} and MambaIRv2 \cite{mambairv22025}, pursue long-range interactions through local enhancement, channel attention and attentive state-space processing.

Efficiency and supervision also shape recent denoising research. SwinIA \cite{swinia2025} uses a self-supervised blind-spot transformer without clean training targets. FCTFANet \cite{fctfanet2025} fuses CNN and transformer features for image restoration, including denoising. However, these methods do not establish performance under our multiplicative gamma-noise protocol in their published results.

Generative restoration methods include DDRM \cite{ddrm2022}, which uses a pretrained diffusion prior for linear inverse problems, and IR-SDE \cite{irsde2023}, which restores images through a learned reverse mean-reverting stochastic differential equation. For SAR despeckling, diffusion with Swin Transformer \cite{rs16173222} and contrastive learning \cite{FANG2024376} specifically address speckle and real-SAR settings. Such results motivate broader comparisons but cannot be transferred directly between additive, SAR and simulated gamma noise. Our experimental extension therefore focuses on controlled evaluation of selected baselines and our proposed approach under gamma noise settings.

\section{Proposed Method: Hybrid++}

\subsection{Multiplicative Gamma Noise Model}
Let $u$ denote the clean grayscale image and $f$ the noisy observation. The multiplicative gamma noise model is:
\begin{equation}
f = u \cdot n, \qquad n \sim \Gamma(L, 1/L),
\end{equation}
where $\Gamma$ is the gamma distribution, $L$ is the number of looks, yielding $\mathbb{E}[n]=1$ and $\text{Var}(n)=1/L$. Low $L$ (e.g., $L{=}1$) corresponds to severe noise, while higher $L$ (e.g., $L{=}10$) represents milder noise levels.

\subsection{Architecture Overview}
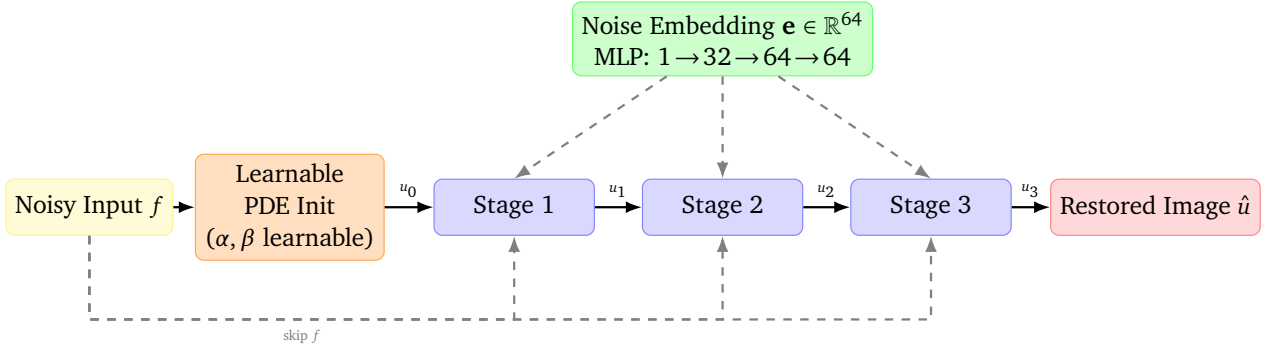
\begin{figure}[!htbp]
\centering
\resizebox{\textwidth}{!}{%
\begin{tikzpicture}[
    box/.style={
        draw, rounded corners=3pt, minimum width=2.0cm, minimum height=0.7cm,
        align=center, font=\small
    },
    pde/.style={box, fill=orange!25, draw=orange!70},
    stage/.style={box, fill=blue!15, draw=blue!50},
    embed/.style={box, fill=green!20, draw=green!60},
    outputbox/.style={box, fill=red!15, draw=red!50},
    arr/.style={-{Latex[length=2mm]}, thick},
    darr/.style={-{Latex[length=2mm]}, thick, dashed, gray},
]

\node[box, fill=yellow!20, draw=yellow!60] (f) at (-2.5, 0) {Noisy Input $f$};
\node[pde]   (pde) at (0.0, 0) {Learnable\\PDE Init\\($\alpha,\beta$ learnable)};
\node[stage] (s1)  at (2.8, 0) {Stage 1};
\node[stage] (s2)  at (5.4, 0) {Stage 2};
\node[stage] (s3)  at (8.0, 0) {Stage 3};
\node[outputbox]   (out) at (10.8, 0) {Restored Image $\hat{u}$};

\draw[arr] (f) -- (pde);
\draw[arr] (pde) -- node[above, font=\tiny]{$u_0$} (s1);
\draw[arr] (s1) -- node[above, font=\tiny]{$u_1$} (s2);
\draw[arr] (s2) -- node[above, font=\tiny]{$u_2$} (s3);
\draw[arr] (s3) -- node[above, font=\tiny]{$u_3$} (out);

\draw[darr] (f) -- ++(0,-1.4) -| (s1.south)
    node[near start, below, font=\tiny]{skip $f$};
\draw[darr] (f) -- ++(0,-1.4) -| (s2.south);
\draw[darr] (f) -- ++(0,-1.4) -| (s3.south);

\node[embed] (emb) at (5.4, 2.1)
    {Noise Embedding $\mathbf{e}\in\mathbb{R}^{64}$\\MLP: $1\!\to\!32\!\to\!64\!\to\!64$};

\draw[darr] (emb) -- (s1.north);
\draw[darr] (emb) -- (s2.north);
\draw[darr] (emb) -- (s3.north);

\end{tikzpicture}%
}

\vspace{2mm}
\caption{Hybrid++ architecture diagram. The noisy image $f$ passes through a differentiable PDE initialization with learnable $\alpha,\beta$ and three enhanced reaction-diffusion stages before the final output. Dashed arrows denote noisy-image skip connections.}
\label{fig:architecture}
\end{figure}

Hybrid++ is built directly upon the Trainable Nonlinear Reaction Diffusion (TNRD) framework \cite{Chen_2017}, preserving its core mathematical ideation - a sequence of reaction-diffusion stages, each comprising learnable convolution filters, a nonlinear influence function and a data-fidelity reaction term while systematically upgrading them for suiting multiplicative gamma noise and giving better performance. Hybrid++ is not a purely hand-crafted PDE with all parameters fixed analytically. Rather, the denoising trajectory remains organized as explicit reaction-diffusion updates. The PDE initialization, multiscale diffusion filters, bounded influence function, stage-wise fidelity weights $\lambda_t$, and noise-conditioned reaction term each have identifiable roles in the restoration dynamics. Thus, Hybrid++ demonstrates structured interpretability. Table~\ref{tab:tnrd_upgrade} shows the component-wise enhancements and Fig~\ref{fig:architecture} shows the network architecture.

\begin{table}[!htbp]
\caption{TNRD $\to$ Hybrid++: Component-wise differences}
\vspace{2mm}
\label{tab:tnrd_upgrade}
\centering
\small
\setlength{\tabcolsep}{4pt}
\renewcommand{\arraystretch}{1.1}
\begin{tabular}{p{2.3cm} p{3.5cm} p{5.5cm}}
\toprule
Component & TNRD & Hybrid++ \\
\midrule
Initialization & Identity (noisy) & Learnable PDE \\
Filter input & 1-channel & 2-channels \\
Filter bank & 8 single-scale ($3{\times}3$) & 64 multiscale ($3{\times}3$ + $5{\times}5$) \\
Influence fn. & 1-layer per-ch $\tanh$ & 4-layer with SE attention \\
Noise adapt. & None (separate models) & 64-dim MLP embedding \\
Stages & 7 & 3 \\
\bottomrule
\end{tabular}
\end{table}

The architecture consists of two principal components: (1) a \textit{learnable PDE initialization} that replaces TNRD's noisy initialization with a PDE-based prior, and (2) a sequence of 3 \textit{enhanced reaction-diffusion stages} that iteratively refine this estimate using the upgraded components described below.

\subsection{Differentiable PDE Initialization}
The process starts with a nonlinear smooth diffusion PDE \cite{perona1990}:
\begin{equation}
\frac{\partial u}{\partial t} = \operatorname{div}\!\left( \left(\frac{u_\sigma}{M}\right)^{\!\alpha} \frac{1}{1+|\nabla u_\sigma|^\beta}\, \nabla u \right),
\label{eq:pde}
\end{equation}
where $u_\sigma$ denotes the Gaussian-smoothed estimate, $M$ is its spatial maximum intensity, and $\alpha$, $\beta$ control the intensity and gradient dependent diffusion behaviors respectively.

\textbf{Key enhancement:} Unlike classical PDE \cite{perona1990} where $\alpha$ and $\beta$ are fixed scalars decided by grid search, Hybrid++ makes these learnable, allowing the model to jointly optimize the PDE alongside the convolutional stages using backpropagation. This enables the PDE initialization to adapt its diffusion behavior to the specific noise it is dealing with during training.

\subsection{Multiscale Filter Banks with Skip Connections}
At each stage $t$, the network receives a 2-channel input formed by concatenating the current estimate $u_{t-1}$ with the original noisy observation $f$. This skip-connection \cite{he2016resnet} provides access to high-frequency information that otherwise would have been lost through successive diffusion steps.

This input is processed by two parallel and independent learnable filter banks:
\begin{enumerate}
\item \textbf{32 filters of size $3{\times}3{\times}2$}: A $3{\times}3$ kernel operating on the 2 channel input (the concatenation of the current estimate $u_{t-1}$ and the original noisy image $f$) to capture fine-grained local textures and sharp edge transitions.
\item \textbf{32 filters of size $5{\times}5{\times}2$}: A  $5{\times}5$ kernel to capture broader spatial context and large structural patterns.
\end{enumerate}

The filter responses are concatenated along the channel dimension, producing a 64-channel feature map. This design is essential for multiplicative noise where noise patterns overlap a lot with structural high-frequency information. The dual-scale representation allows the network to separate noise from texture at multiple spatial frequencies simultaneously. All filters are initialized with Kaiming-style scaling \cite{he2015delvingdeeprectifierssurpassing} and are fully learnable throughout training.

\subsection{Dynamic Noise Level Embedding}
To enable noise-conditioned denoising across varying known noise levels, a 3-layer Multi-Layer Perceptron (MLP) \cite{popescu2009mlp} maps the scalar noise strength parameter to a 64-dimensional conditioning vector:
\begin{equation}
\mathbf{e} = W_3 \cdot \text{ReLU}(W_2 \cdot \text{ReLU}(W_1 \cdot \sigma_L + b_1) + b_2) + b_3,
\end{equation}
where $\sigma_L = 1/\sqrt{L}$ is the noise strength derived from the gamma distribution's coefficient of variation, $b_i$'s are biases and $W_1 \in \mathbb{R}^{32\times1}$, $W_2 \in \mathbb{R}^{64\times32}$, $W_3 \in \mathbb{R}^{64\times64}$.

This 64-dimensional embedding $\mathbf{e}$ is broadcast and added to the 64-channel filter responses at \textit{every} stage. This makes the feature representation biased based on noise level. This design allows Hybrid++ to use a single unified noise-conditioned model for both $L{=}1$ and $L{=}10$. In contrast, the TNRD and DnCNN baselines in our experiments are trained separately for each noise level. Thus, Hybrid++ is evaluated in the more flexible setting: it must learn one conditional mapping across noise severities rather than a separate optimum for each L.

\subsection{4-Layer SE-Attention Influence Function}
A very important component of the reaction-diffusion update is the nonlinear influence function $\Phi(\cdot)$, which decides the diffusion behavior. Hybrid++ introduces a deep, 4-layer architecture with integrated channel attention:

\begin{enumerate}
\item Layer 1: $1{\times}1$ convolution ($64 \to 64$) followed by $\tanh$ activation.
\item Layer 2: $1{\times}1$ convolution ($64 \to 64$) followed by $\tanh$ activation.
\item Layer 3: Squeeze-and-Excitation (SE) block with bottleneck ratio 4:
\begin{equation}
\text{SE}(\mathbf{x}) = \mathbf{x} \cdot \sigma'\!\left(W_{\text{up}} \cdot \text{ReLU}(W_{\text{down}} \cdot \text{GAP}(\mathbf{x}))\right),
\end{equation}
where $W_{\text{down}} \in \mathbb{R}^{16\times64}$ squeezes and $W_{\text{up}} \in \mathbb{R}^{64\times16}$ excites, with $\text{GAP}$ denoting global average pooling and $\sigma'$ the sigmoid function. This mechanism dynamically re-weights the 64 channels based on their global importance, acting as a learned attention mechanism.
\item Layer 4: $1{\times}1$ convolution ($64 \to 1$) followed by $\tanh$, collapsing the attended features into a single spatial diffusion map.
\end{enumerate}

\textbf{Design choice:} The TNRD \cite{Chen_2017} baseline uses a shallow per-channel influence function. By expanding this to 4 layers with an intermediate SE-attention \cite{hu2019squeezeandexcitationnetworks} block, Hybrid++ can learn highly complex nonlinear response mappings. The $\tanh$ activations ensure bounded outputs to stay consistent with the physical diffusion formulation.
Three differences emerge relative to the TNRD baseline:

\begin{enumerate}
    \item \textbf{Cross-channel interaction.}  The SE block \cite{hu2019squeezeandexcitationnetworks} computes a global channel descriptor via average pooling and uses it to re-weight each of the 64 channels based on their global importance. This permits joint selection of filter responses.

    \item \textbf{Input-dependent asymmetric shaping.} Because the SE channel weights depend on the global average of the current feature map, the effective mapping of $\Phi$ changes across input images. The same channel weights are applied at all spatial positions within an image.
    \item \textbf{Bounded nonlinear mapping.} Hybrid++ combines three $\tanh$ activations with an intermediate SE block, providing a bounded final diffusion response.
\end{enumerate}

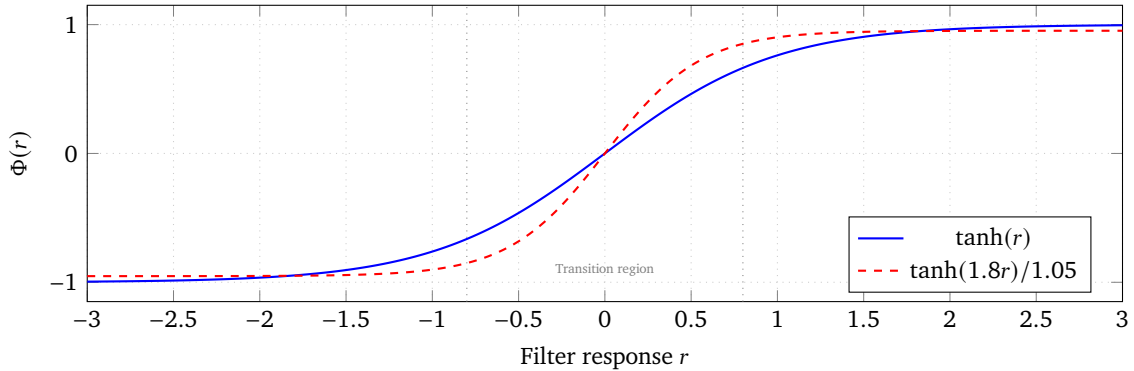
\begin{figure}[!htbp]
\centering
\begin{tikzpicture}
\begin{axis}[
    width=0.92\linewidth, height=5.5cm,
    xlabel={Filter response $r$},
    ylabel={$\Phi(r)$},
    xmin=-3, xmax=3, ymin=-1.15, ymax=1.15,
    legend pos=south east,
    grid=major, grid style={dotted, gray!40},
    tick label style={font=\small},
    label style={font=\small},
    legend style={font=\small},
]
\addplot[blue, thick, domain=-3:3, samples=200]{tanh(x)};
\addlegendentry{$\tanh(r)$}

\addplot[red, thick, dashed, domain=-3:3, samples=200]{tanh(1.8*x)/1.05};
\addlegendentry{$\tanh(1.8r)/1.05$}

\draw[gray, dotted] (axis cs:-0.8,-1.15) -- (axis cs:-0.8,1.15);
\draw[gray, dotted] (axis cs: 0.8,-1.15) -- (axis cs: 0.8,1.15);
\node[font=\tiny, gray] at (axis cs:0, -0.9) {Transition region};
\end{axis}
\end{tikzpicture}
\vspace{2mm}
\caption{%
    Schematic bounded scalar responses: $\tanh(r)$ (solid) and $\tanh(1.8r)/1.05$ (dashed). These analytic curves illustrate response shapes rather than measured TNRD or Hybrid++ functions.
}
\label{fig:phi_comparison}
\end{figure}

Fig.~\ref{fig:phi_comparison} provides a schematic illustration of bounded scalar response shapes. The actual $\Phi$ is a learned multichannel, image-dependent mapping, so these curves are not extracted learned influence functions.

\subsection{Stage Update Rule}
Each stage combines the influenced diffusion term with a data-fidelity reaction term:
\begin{equation}
u_t = u_{t-1} - w \cdot \Phi\!\left(\mathbf{R}(u_{t-1}, f) + \mathbf{e}\right) - \lambda_t \cdot w \cdot \frac{u_{t-1} - f}{u_{t-1}^2 + \varepsilon},
\end{equation}
where $\mathbf{R}(u_{t-1}, f)$ denotes the 64-channel multiscale filter response, $\mathbf{e}$ is the noise embedding, $w$ is the intensity-dependent weight derived from the Gaussian-smoothed estimate and $\lambda_t$ is a learnable fidelity parameter. The reaction term derives from the MAP formulation for multiplicative gamma noise \cite{aubert2008variational}.

\subsection{Two-Phase Training Strategy}
\label{sec:training}
We use a  two-phase optimization strategy:

\textbf{Phase 1: Stage-wise Initialization.} Each of the 3 stages is trained sequentially for 100 epochs. When training stage $t$, all prior stages are frozen and only stage $t$'s parameters alongwith the PDE \cite{perona1990} and embedding parameters are changed. This ensures each stage provides a meaningful refinement step. Training uses Adam with $\text{lr}=10^{-3}$ (learning rate), early stopping with patience 10 and gradient clipping at max norm 1.0.

\textbf{Phase 2: Joint End-to-End Refinement.} After stage-wise initialization, \textit{all} parameters across the 3 stages, the PDE module \cite{perona1990} and the noise embedding are unfrozen and jointly optimized for 50 epochs. This phase uses Adam with $\text{lr}=10^{-4}$ and CosineAnnealingLR \cite{loshchilov2017sgdrstochasticgradientdescent} ($T_{\max}{=}50$, $\eta_{\min}{=}10^{-6}$), allowing the entire network to co-adapt and fine-tune the inter-stage interactions. We use Mean Squared Error (MSE) as the loss function for both phases. 

The purpose of the two-phase strategy is not to keep the stages independent after training, but to provide a structured initialization before global co-adaptation. The stage-wise phase encourages each reaction-diffusion block to learn a meaningful refinement role. The later joint phase then fine-tunes the already structured initialization across the PDE module, noise embedding and all diffusion stages. Therefore, joint training refines an initialization in which the stages have already learned progressively useful denoising roles.

\section{Experimental Setup}
\subsection{Dataset}
We use the FoE dataset \cite{roth2005foe} containing 400 grayscale images that are split into 80\% training, 10\% validation, and 10\% testing (40 test images) with a fixed random seed of 42. Images have size $180\times180$ and grayscale intensities normalized to $[0,1]$. For qualitative evaluation on real-world data, we additionally show real SAR examples from the Sentinel-1\&2 dataset; these lack clean references.

\subsection{Baselines}
We compare Hybrid++ against five models spanning classical and learned denoising:
\begin{enumerate}
\item \textbf{BM3D} \cite{dabov2007bm3d}: Non-local patch-based method applied using homomorphic (log-domain) transform with noise variance $\sigma^2 = \psi_1(L)$ (trigamma function).
\item \textbf{PDE Baseline} \cite{perona1990}: Classical nonlinear smooth diffusion PDE (Eq.~\ref{eq:pde}) with tuned parameters ($\alpha{=}1.0$, $\beta{=}2.0$).
\item \textbf{TNRD} \cite{Chen_2017}: The trainable reaction-diffusion baseline. We trained separate models for each noise level.
\item \textbf{DnCNN} \cite{Zhang_2017}: 17-layer deep CNN with batch normalization (approx. 556K parameters) for the black-box deep neural network architecture.
\item \textbf{Uformer-T} \cite{Wang_2022_Uformer}: The tiny U-shaped transformer, reproduced for grayscale gamma denoising and described in Section~\ref{sec:uformer}.
\end{enumerate}
The baselines cover classical PDE priors, non-local denoising, trainable reaction--diffusion, CNNs and a transformer. Our TNRD baseline is an original TNRD-inspired re-implementation with seven stages and eight fixed, randomly initialized \(3\times3\) filters per stage. It learns per-filter affine–tanh influence functions, stage-specific fidelity weights and skip-connection gates, totaling 126 trainable parameters. Learning-based models use MSE loss and Adam with the same 80/10/10 split specification.

\textbf{Remark on log-domain vs.\ direct-domain processing.}
BM3D \cite{dabov2007bm3d} is an additive-noise denoiser.  To handle multiplicative gamma noise we apply it via a homomorphic transform: the log of the noisy image is computed, BM3D is applied with standard deviation
$\sigma = \sqrt{\psi_1(L)}$ (where $\psi_1$ is the trigamma function, the log-domain noise variance of a $\Gamma(L,1/L)$ variate), and the result is exponentiated. All other models - PDE, TNRD, DnCNN,  Uformer-T and Hybrid++ are applied directly in the original image domain without any log transformation.

\subsection{Metrics and Reproducibility}
PSNR measures pixel fidelity, $\mathrm{PSNR}=10\log_{10}(1/\mathrm{MSE})$ in dB for unit-range images; SSIM measures local luminance, contrast and structural agreement \cite{wang2004ssim}. Scores are averaged per image. SSIM uses its default $7\times7$ uniform window. No border pixels are discarded.

The official code and reproducibility details will be available on GitHub at: \textit{github.com/Sujato-Dutta/Hybrid-Image-Denoising-for-Gamma-Noise-Removal}.

\section{Results and Discussion}

\subsection{Quantitative Results}
Table~\ref{tab:main_results} reports the average PSNR and SSIM \cite{wang2004ssim} across all 40 test images for comparison of the six models.

\begin{table}[t]
\centering
\caption{FoE comparison at $L=1$ and $L=10$}
\label{tab:main_results}
\begin{tabular}{lrrrrr}
\toprule
Method & Params & PSNR$_1$ & SSIM$_1$ & PSNR$_{10}$ & SSIM$_{10}$ \\
\midrule
BM3D & 0 & 12.67 & 0.384 & 24.64 & 0.755 \\
PDE Baseline & 0 & 17.66 & 0.422 & 23.07 & 0.651 \\
TNRD & 126 & 18.55 & 0.324 & 23.17 & 0.628 \\
DnCNN & 556,097 & 21.09 & 0.484 & 25.94 & 0.761 \\
Uformer-T & 5,290,987 & 21.97 & 0.540 & 26.65 & 0.776 \\
\textbf{Hybrid++(Ours)} & \textbf{44,408} & \textbf{20.86} & \textbf{0.463} & \textbf{25.02} & \textbf{0.692} \\
\bottomrule
\end{tabular}
\par\smallskip\footnotesize
Proposed method is bolded.
\end{table}

Under the severe noise regime ($L{=}1$), Hybrid++ achieves \textbf{20.86~dB}, surpassing the PDE baseline \cite{perona1990} by \textbf{+3.20~dB}, TNRD \cite{Chen_2017} by \textbf{+2.31~dB} and BM3D \cite{dabov2007bm3d} by \textbf{+8.19~dB}. At the milder setting ($L{=}10$), Hybrid++ leads with \textbf{25.02~dB} dominating all the interpretable baselines by a considerable margin. For SSIM, it stays clearly better for the ($L{=}1$) severe noise setting while trailing behind BM3D \cite{dabov2007bm3d} by just \textbf{0.063} at ($L{=}10$). \newline When compared to the black-box model DnCNN  \cite{Zhang_2017}, it comes close at severe noise level ($L{=}1$) trailing by just \textbf{0.23~dB} PSNR and \textbf{0.021 SSIM}. For ($L{=}10$), the gap is wider at \textbf{0.92~dB} PSNR and \textbf{0.069} SSIM. However, Hybrid++ achieves this using one unified noise-conditioned model, while DnCNN is trained separately for each L. It also uses only \textbf{44k} parameters as compared to DnCNN's \textbf{556k} while maintaining structured interpretability vs DnCNN's black-box architecture. The comparison with Uformer has been done in details in a separate section below.

\subsection{Qualitative Analysis}

\begin{figure}[!htbp]
\centering
\includegraphics[width=0.8\textwidth]{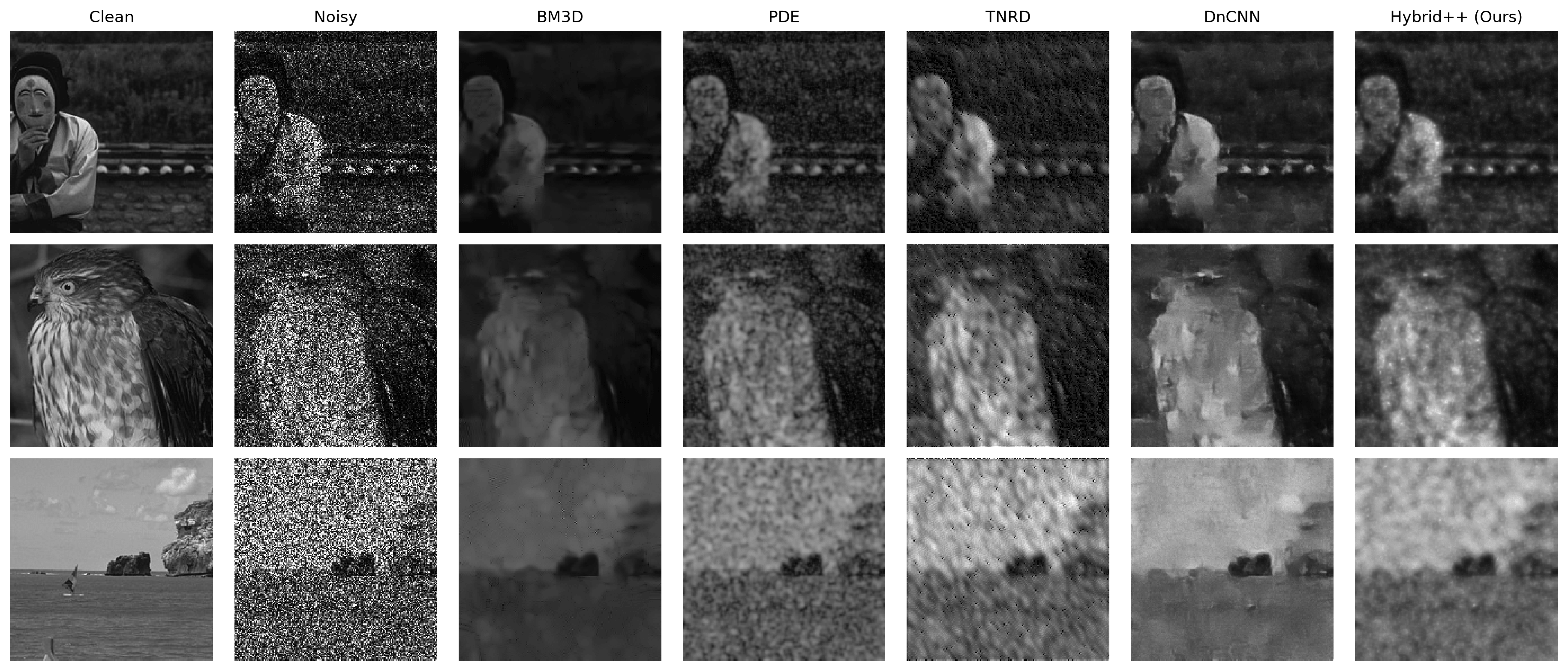}
\vspace{2mm}
\caption{Qualitative comparison on FoE test images under severe gamma noise ($L{=}1$).}
\label{fig:qual_l1}
\end{figure}

\begin{figure}[!htbp]
\centering
\includegraphics[width=0.8\textwidth]{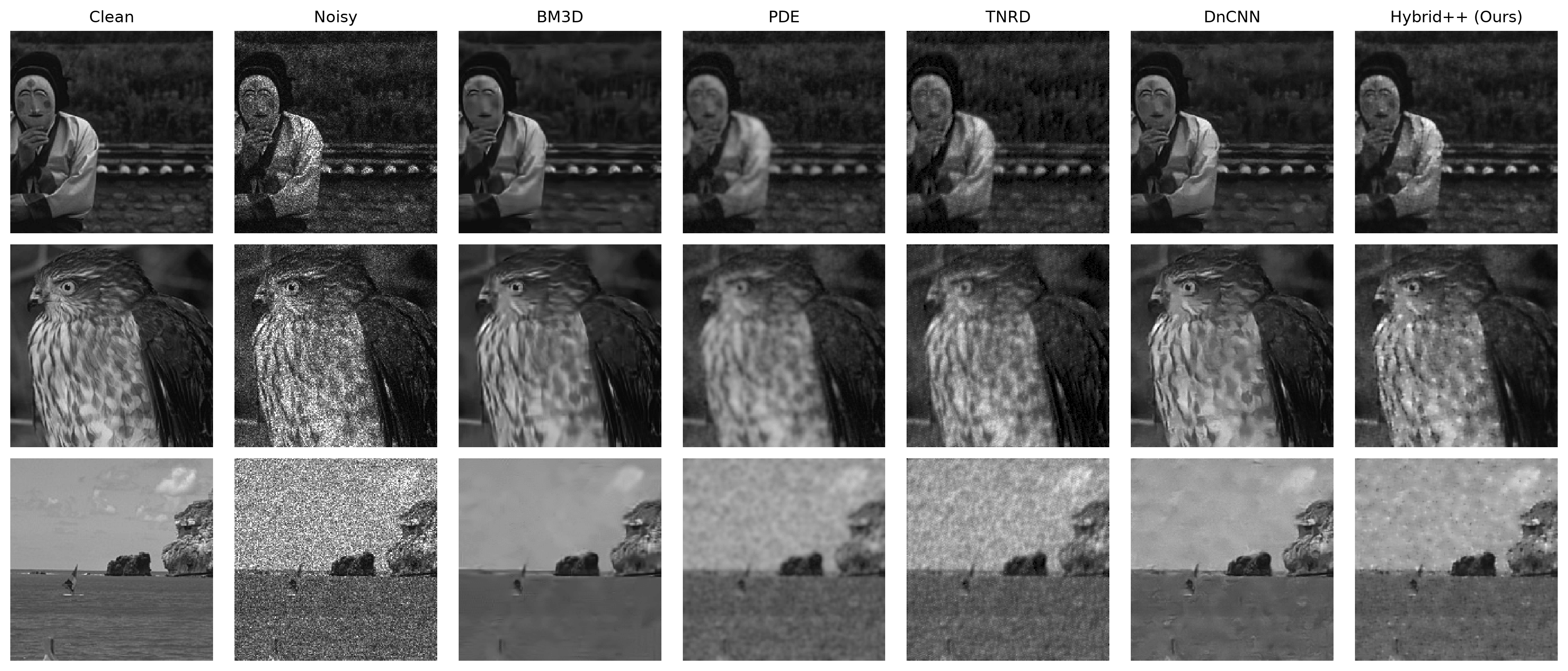}
\vspace{2mm}
\caption{Qualitative comparison on FoE test images under mild gamma noise ($L{=}10$).}
\label{fig:qual_l10}
\end{figure}

Figs.~\ref{fig:qual_l1} and~\ref{fig:qual_l10} show us the qualitative comparisons under $L{=}1$ and $L{=}10$ noise levels respectively. Under the severe $L{=}1$ regime, BM3D \cite{dabov2007bm3d} fails due to the inadequacy of its homomorphic log-domain adaptation for extreme multiplicative noise. PDE \cite{perona1990} performs over-smoothing, thus losing structural information. TNRD \cite{Chen_2017} does comparatively better but it shows blurring of the edges and haziness over the image. Hybrid++ recovers significantly sharper edges and more coherent textures approaching the performance of DnCNN \cite{Zhang_2017}. Under the milder $L{=}10$ setting, the differences between methods narrow, but Hybrid++ consistently preserves finer granular detail that PDE and TNRD diffuse away.

\subsection{Bridging the Gap to Deep Learning}
DnCNN, with approximately 556K parameters, achieves 21.09 dB at $L{=}1$. Hybrid++, with 44,408 parameters, reaches 20.86 dB, leaving a gap of only 0.23 dB while using roughly $12.5\times$ fewer parameters. This serves as an evidence that a compact physically structured model can approach a representative black-box CNN under severe multiplicative noise. At $L{=}10$, DnCNN remains ahead, especially in SSIM. Hybrid++ trades some peak restoration performance for parameter compactness, unified noise conditioning and structured reaction-diffusion interpretability.

\subsection{Training Convergence Analysis}

\begin{figure}[!htbp]
\centering
\includegraphics[width=0.7\linewidth]{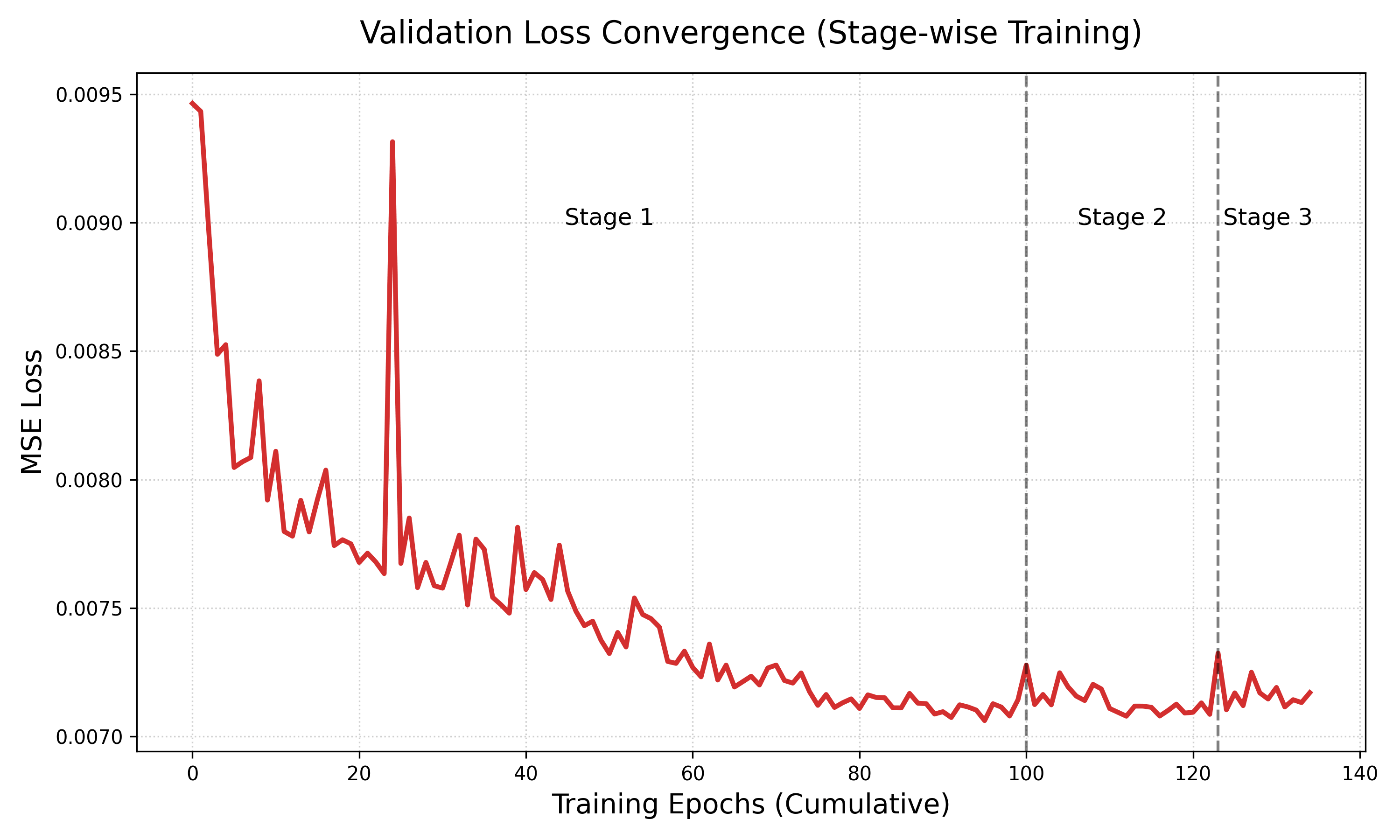}
\vspace{2mm}
\caption{Validation loss during stage-wise training. Stage 1 achieves the dominant loss reduction, with diminishing returns in subsequent stages of the tested three-stage model.}
\label{fig:convergence}
\end{figure}

Fig.~\ref{fig:convergence} illustrates the validation loss trajectory over the stage-wise training phase. Stage~1 shows a sharp decline in loss as the model learns the primary denoising transformation. Stage~2 provides a modest further reduction, refining the residual details. Stage~3 shows near-flat convergence, indicating diminishing returns within the tested three-stage model. This curve does not establish that deeper models cannot improve performance. This convergence behavior describes the stage-wise initialization used before joint refinement. The stage-wise phase gives the stages distinct denoising roles before the final joint refinement: Stage 1 performs the dominant restoration, Stage 2 provides residual correction, and Stage 3 performs only a small final adjustment. The later end-to-end phase therefore fine-tunes an already structured sequence rather than learning the full model from an unstructured initialization.

\subsection{Qualitative Evaluation on Real SAR Images}

\begin{figure}[!htbp]
\centering
\includegraphics[width=0.8\linewidth]{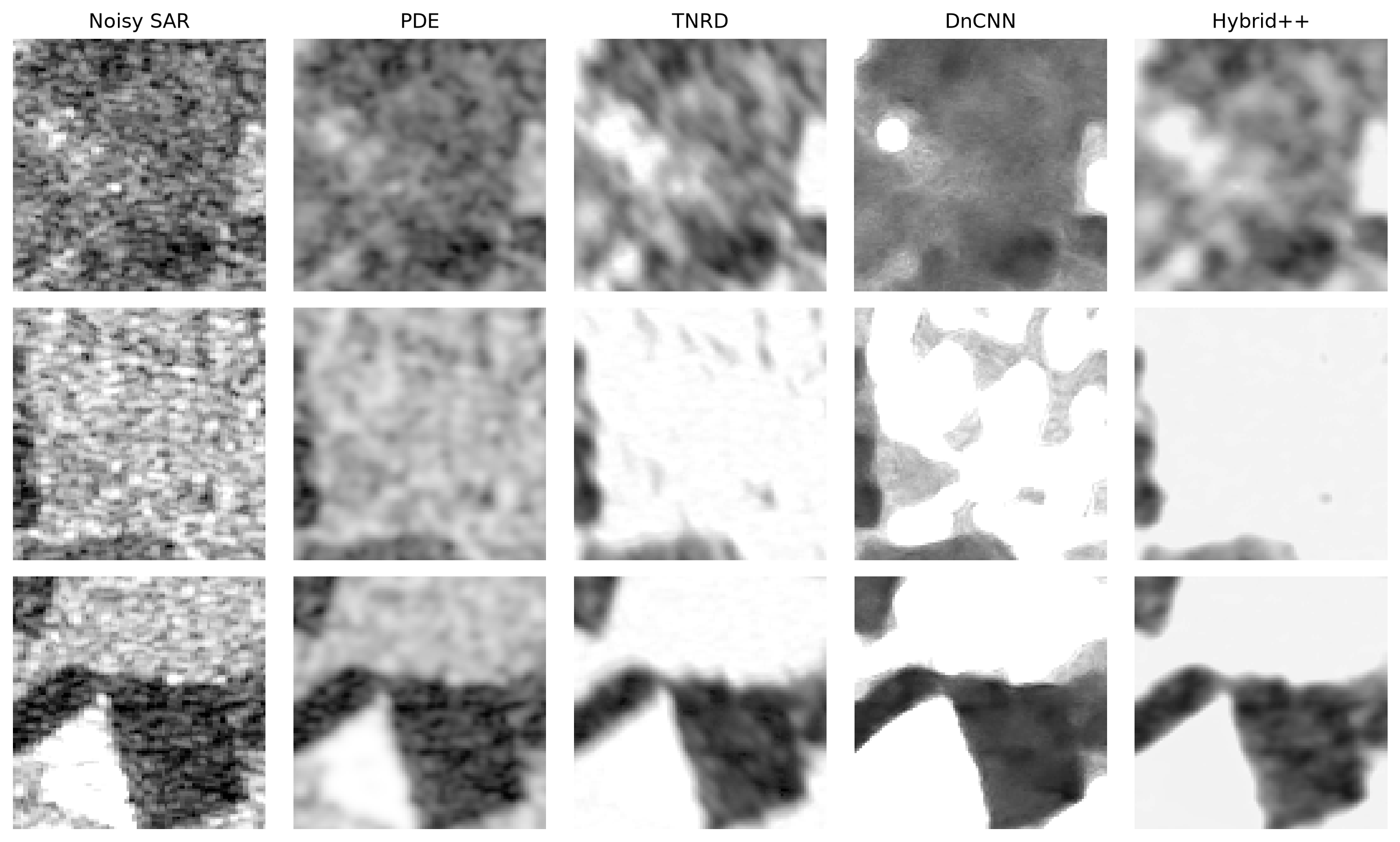}
\vspace{2mm}
\caption{Qualitative comparison on real SAR images (Sentinel 1\&2 dataset). Hybrid++ is shown last. The lower two rows exhibit saturation and detail loss, illustrating a failure mode under real-SAR domain shift.}
\label{fig:sar}
\end{figure}

Fig.~\ref{fig:sar} presents denoised results on real SAR examples. Since there is no clean ground truth, this comparison is qualitative and does not establish a quantitative ranking. Although broad structures remain visible, the lower two rows show excessive smoothing and near-saturated regions in Hybrid++ outputs. These examples reveal a limitation of transferring a model trained on simulated gamma corruption to real SAR imagery.

\section{Comparison with Transformer-based architecture}
\label{sec:uformer}
\subsection{Uformer Baseline}
Uformer is a transformer-based restoration architecture \cite{Wang_2022_Uformer}. We evaluate Uformer-T with embedding dimension 16 and two blocks at each of nine levels across the encoder, bottleneck and decoder. It uses $8\times8$ window attention, locally enhanced feed-forward layers and decoder modulation. Separate models for $L=1$ and $L=10$ are trained for up to 100 epochs, batch size 4, using Adam at $10^{-4}$. The learning rate is halved after five validation epochs without improvement; early-stopping patience is 10 and gradient-norm clipping is 1.

\subsection{Results and Discussion}
Uformer-T obtains 21.97/26.65 dB PSNR and 0.540/0.776 SSIM at $L=1/10$ (Table~\ref{tab:main_results}), exceeding the Hybrid++ scores by 1.11/1.63 dB and 0.077/0.084 SSIM.
The main distinction is the structure of the denoising mapping. Hybrid++ represents restoration as three explicit reaction--diffusion updates which have identifiable mathematical roles. They allow the update trajectory to be inspected in terms of smoothing and fidelity to the observation. Uformer-T instead learns attention and feature transformations throughout a multiscale hierarchy; its latent features do not directly correspond to a specified gamma-noise diffusion equation. Attention weights can be inspected, but this does not by itself provide the same physical interpretation of each update. Hybrid++ therefore offers structured interpretability using 44,408 parameters against Uformer-T's 5,290,987, approximately $119\times$ fewer parameters. It also requires fewer counted MACs in our measurements (Section~\ref{sec:complexity}).

\section{Complexity Analysis}
\label{sec:complexity}
Table~\ref{tab:complexity} reports batch-one FP32 inference on an NVIDIA A100-SXM4-40GB MIG device using PyTorch 2.8.0, CUDA 12.8 and four CPU threads with TF32 disabled. Median latency follows 20 warm-up and 100 synchronized timed passes. MAC counts include convolution, linear and attention matrix products; nonlinearities, normalization and reductions are excluded. One MAC corresponds to two arithmetic FLOPs. Uformer padding is included in the measured path.

\begin{table}[t]
\centering
\caption{Architecture inference at batch 1, FP32 on NVIDIA A100-SXM4-40GB MIG. MACs count convolution, linear and attention matrix products; nonlinear operations excluded.}
\label{tab:complexity}
\begin{tabular}{lrrrr}
\toprule
Method & Size & Params & GMAC & Median ms \\
\midrule
PDE baseline & 180 & 0 & 0.016 & 13.14 \\
PDE baseline & 256 & 0 & 0.033 & 13.13 \\
Hybrid++ & 180 & 44,408 & 1.033 & 15.55 \\
Hybrid++ & 256 & 44,408 & 2.089 & 17.78 \\
DnCNN & 180 & 556,097 & 17.953 & 6.83 \\
DnCNN & 256 & 556,097 & 36.314 & 13.44 \\
TNRD & 180 & 126 & 0.040 & 2.35 \\
TNRD & 256 & 126 & 0.081 & 2.37 \\
Uformer-T & 180 & 5,290,987 & 11.212 & 36.14 \\
Uformer-T & 256 & 5,290,987 & 11.212 & 36.12 \\
\bottomrule
\end{tabular}
\end{table}

At $180\times180$, Hybrid++ uses 1.033 GMAC versus Uformer-T's 11.212 GMAC and has median latency 15.55 ms versus 36.14 ms. However, DnCNN is faster at 6.83 ms despite more parameters and MACs; sequential PDE updates and operator execution overhead prevent a universal latency advantage.

\section{Ablation Study}
To justify our architectural design choices in Hybrid++, we conducted an ablation study by modifying the base TNRD \cite{Chen_2017} configuration and comparing PDE initialization variants.

\begin{table}[!htbp]
\caption{Ablation Study: Impact of Different Components}
\vspace{2mm}
\label{tab:ablation}
\centering
\begin{tabular}{lcccc}
\toprule
& \multicolumn{2}{c}{$L=1$} & \multicolumn{2}{c}{$L=10$} \\
\cmidrule(lr){2-3} \cmidrule(lr){4-5}
Variant & PSNR & SSIM & PSNR & SSIM \\
\midrule
Full Base TNRD & 18.55 & 0.324 & 23.17 & 0.628 \\
\quad $-$ Nonlinear influence & 10.46 & 0.144 & 17.63 & 0.456 \\
\quad $-$ Reduced filters ($N_k{=}4$) & 16.91 & 0.245 & 21.99 & 0.559 \\
\quad $-$ Reduced stages ($T{=}3$) & 17.06 & 0.289 & 22.24 & 0.578 \\
PDE Baseline & 17.65 & 0.421 & 23.07 & 0.651 \\
PDE initialization + Base TNRD & 20.09 & 0.411 & 23.99 & 0.680\\
\midrule
\textbf{Hybrid++ (Full)} & \textbf{20.86} & \textbf{0.463} & \textbf{25.02} & \textbf{0.692} \\
\bottomrule
\end{tabular}
\end{table}

Table~\ref{tab:ablation} reveals the importance and impact of the various design choices we made for our final architecture:

\textbf{Nonlinear influence functions -} $\phi$ is replaced by the identity function ($\phi(r) = r$), reducing the influence function to a plain linear filter response. This shows a catastrophic ${\sim}8$~dB drop at $L{=}1$, reducing performance on severe noise levels. This supports the importance of nonlinear influence functions in the tested baseline configuration.

\textbf{PDE initialization -} The PDE initialization provides a strong starting point before the learned diffusion stages are applied. As shown in Table~\ref{tab:ablation}, the PDE baseline already achieves competitive structural preservation. When combined with base TNRD \cite{Chen_2017}, it improves performance from 18.55/23.17 dB PSNR to 20.09/23.99 dB while also boosting SSIM.

\textbf{Filter capacity -} Reducing filters from 8 to 4 drops PSNR by 1.64~dB at $L{=}1$. This motivates exploring greater filter capacity. 

\textbf{Stage depth -} Reducing from 7 to 3 stages drops PSNR by 1.49~dB in the base TNRD (with 8 filters). However, Hybrid++ demonstrates that with sufficient filter capacity (64 channels) and deep influence functions, a three-stage model achieves the reported accuracy.

\section{Limitations and Failure Analysis}
\label{sec:limitations}
Quantitative evaluation is limited to 40 held-out FoE images with simulated gamma noise at $L=1$ and $L=10$. SAR examples lack clean references and quantitative reference-free validation. 

\begin{figure}[!htbp]
\centering
\includegraphics[width=0.7\linewidth]{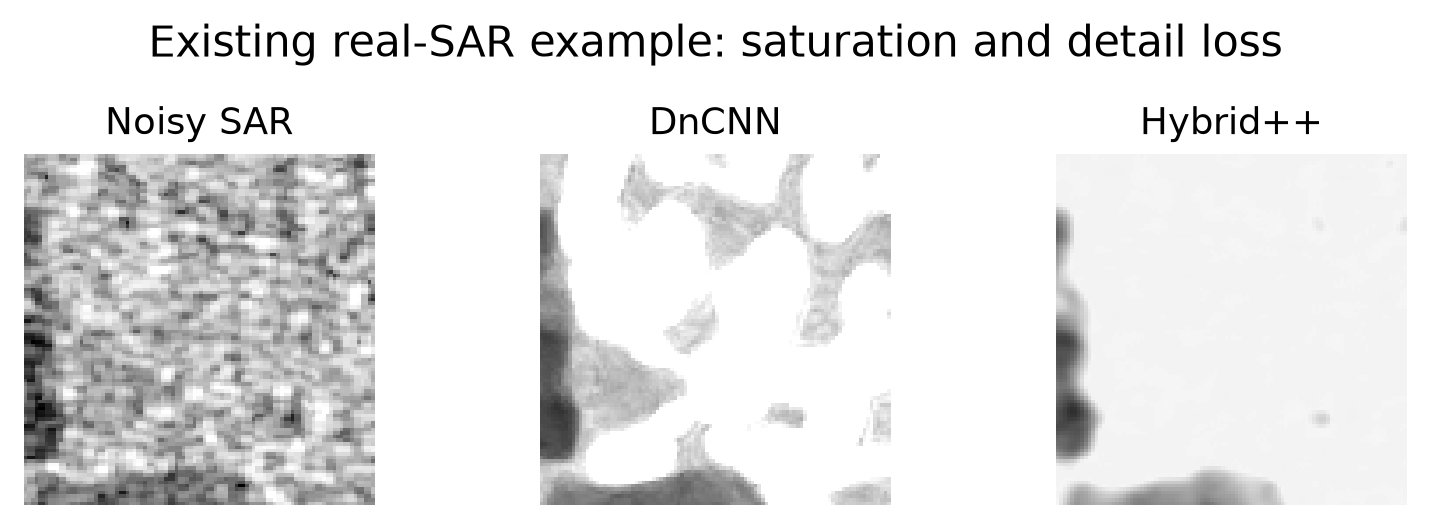}
\caption{Failure example from SAR comparison: noisy observation, DnCNN and Hybrid++. The Hybrid++ output exhibits saturation and loss of fine detail.}
\label{fig:failure}
\end{figure}

The examples shown in Fig.~\ref{fig:failure} show loss of fine details due to oversmoothing in Hybrid++. No clean SAR reference is available to quantify the lost detail. More generally, low-contrast textures and thin structures can be confused with noise and smoothed away, while intense reflectors and out-of-distribution contrast can produce clipped or distorted reconstructions. These are plausible failure settings motivated by the model assumptions and the displayed examples. Broader real-noise training and evaluation are needed before extending the method to operational SAR or medical imaging.

\section{Conclusion}
This paper presented Hybrid++, a physics-informed architecture for multiplicative gamma noise removal that bridges the divide between classical hand-crafted PDE-based methods and modern deep learning. By expanding the TNRD framework with learnable PDE initialization, 64-channel multiscale filter banks, 4-layer SE-attention influence functions, 64-dimensional noise embeddings and a two-phase training strategy, Hybrid++ achieved 20.86~dB at $L{=}1$ and 25.02~dB at $L{=}10$ on the FoE benchmark while only requiring to train a single unified model. Most significantly, a 44,408-parameter model closed the gap to the 556K-parameter DnCNN to within just 0.23~dB at severe noise levels. These results suggest the effectivity of principled physical priors when augmented with targeted learnable components in the severe gamma-noise regime. Future work should extend evaluation to additional restoration models, broader SAR datasets and real-noise training protocols.
%
%
%
\bibliographystyle{plainnat}
\bibliography{bibliography}
\end{document}